# Identifying downregulated hub genes and key pathways in HBV-related hepatocellular carcinoma using systems biology approach

**Dr. Sedigheh Behrouzifar**

Department of Medical Sciences, Shahrood Branch,

Islamic Azad University, Shahrood, Iran

sedighehbehrouzifar@gmail.com

**Abstract**

*Chronic Hepatitis B (CHB) is an independent risk factor for hepatocellular carcinoma (HCC) initiation without cirrhosis occurrence. Apart from the favorable effects of some antiviral drugs following tumor resection on the survival of HCC patients, the use of these agents is essential lifelong. Thus, designing the target-oriented therapeutic strategies to increase life expectancy in HCC patients would be very important. The present study aimed to identify downregulated hub genes and enriched pathways in HB-related HCC using a systems biology-based approach. Microarray data of GSE121248 were downloaded from gene expression omnibus (GEO) database. The differentially expressed genes (DEGs) with the cut-off criteria of adjusted $p < 0.05$ and Log Fold-change (FC) $\leq$ -1.5 were selected. Then, the genes with the highest centrality were detected. Finally, the prognostic values of the hub genes were assessed. Six under-expressed hub genes with the highest interaction degree, Betweenness and Eigenvector centrality were including IGF-1, PTGS2, PLG, HGF, ESR-1 and CYP2B6. Among genes with high centrality, several genes including CYP2C9, ESR-1, CXCL2, CYP2C8, IGF-1, CYP3A4, CYP2E1, CERPINE-1 and PXR were prognostic in HCC. The important repressed pathways were including metabolic pathways and PI3K-Akt and chemokine signaling pathways. The under-expression of several genes implicated in metabolism, differentiation and chemotaxis might be a hallmark of the progression of HCC that can be*

*considered as diagnostic and therapeutic targets.*

1- Introduction

Liver cancer is the fourth leading cause of cancer death worldwide, with 841,000 incidence rates and 782,000 deaths annually. Almost 80 % of cases of liver cancer are associated with HCC. The incidence of HCC in females before menopause is far lower than that in males due to protective effects of estrogen [1]. As many as 25% of HBV-infected patients will develop HCC. Since 1982, primary prevention against HCC originating from HB has been feasible through the hepatitis B vaccine [2]. According to estimates from the World Health Organization in 2018, at least 60% of liver cancer cases are due to late diagnosis and treatment of viral hepatitis.

The integration of the hepatitis B virus DNA into genome of the hepatocytes of chronic HBV carriers leads to producing HBV X protein (HBx). Carcinogenesis role of HBx protein has been demonstrated in the previous literature [3]. HBx alters the expression of the genes involved in cell cycle, transcriptional regulation, immune response, and metabolism [4]. High expression of pro-inflammatory factors such as chemokines and cytokines in liver by HBx results in peripheral leukocyte infiltration to liver [5], chronic necro-inflammation of hepatocytes, liver abnormal regeneration, genomic unstability and conclusively HCC [6]. Moreover, HBx results in occurrence of epigenetic (methylation and de-acetylation) changes and under-expression of tumor-suppressor genes [3, 7].

A large percentage of the HCC patients identified at the advanced stages have poor prognosis, and owing to organ donation shortage, liver transplantation is limited. Advanced HCC is with immunosuppression, and systemic chemotherapy gives rise attenuating defensive system of body. In addition, HBV reactivation during immunosuppression is a critical issue.

Therefore, the scrutiny of the molecular mechanism of HBV carcinogenesis in cancerous hepatocytes and tumor microenvironment is of pivotal importance in the development of novel strategies for optimal management of HCC. Furthermore, the focus of treatment based on prognostic genes is indispensable. In the present study, we analyzed microarray expression profile to explore gene expression variations in HB-related HCC. The objectives of this study were identifying differentially downregulated radical biomarkers (genes), prioritizing hub genes based on the highest centrality, awareness into the behavior of the modules (sub-networks) in protein-protein interaction (PPI) network via ranking the pathways, biological processes and molecular functions enriched by hub interactive genes using systems biology based approach. Finally, we studied the prior literature to validate the role of hub genes in HCC and interpreted the possible effects of dysregulation of hub genes on HCC outcomes. However, additional investigations into how HCC develops from hepatitis B are essential in future researches.

## 2- Material and methods

### 2-1- Microarray data

Human microarray data of GSE121248 deposited into GEO database (http://www.ncbi.nlm.nih.gov/geo/) in 2018 was downloaded. The differentially expressed genes (DEGs) were determined in 45 HBsAg positive cancerous liver tissue samples and 28 HBsAg positive non-cancerous tissue samples using limma R packages (version 3.4.1) in Bioconductor (http://www.bioconductor.org/). Because of hepatoprotective effect of estrogen in females, the tissue samples of male patients were merely used. The gene expression data were normalized using the mas5 algorithm. Gene expression data for individual samples was calculated using gene expression console software.

### 2-2- Identification of DEGs

In the first step, probes (genes) with adjusted $p < 0.05$ were screened. Then, downregulated genes with Log fold change (FC) $\leq$ -1.5 cut-off criteria were selected. Protein-protein interaction network related to the DEGs were obtained via STRING (The Search Tool for the Retrieval of Interacting Genes) online database (available online: http://string-db.org). Then, DEGs were further screened using cytoscape software and 111 hub genes were selected. Pathway enrichment analysis of the hub genes was conducted with STRING database and Enrichr web server (http://amp.pharm. mssm.edu/ Enrichr/). The importance of physiological function of hub genes (Centrality) was computed and visualized using Gephi software. In this step, genes with the highest Degree, Closeness, Betweenness and Eigen vector centrality were detected. Modularity class was determined using Gephi software and annotated using Kyoto Encyclopedia of Genes and Genomes (KEGG), Reactome and Wikipathway enrichment analysis in Enrichr web server. In addition, Gene Ontology (GO) analysis was performed for the assay of biological processes, molecular functions of genes into modules and cellular components where the hub genes are involved. Finally, in order to analyze the overall survival of the HCC patients, the prognostic values of the hub genes with high centrality were calculated using the Kaplan-Meier method through Gepia online software.

## 3- Results

### 3-1- The identification of the differentially expressed genes and hub genes in HCC

At First, 13720 probes (genes) with adjusted $p < 0.05$ were selected. Of these, 510 DEGs with log FC $\leq$ -1.5 were extracted. Using Cytoscape software, 111 under-expressed hub genes were screened. Protein-protein interaction network of the hub genes were represented by STRING database (Supplementary Fig. 1).

### 3-2- Pathway enrichment and GO function analysis

The under-expressed hub genes were enriched in the pathways including metabolic pathways (hsa01100), retinol metabolism (hsa00830), drug metabolism (hsa00982), chemical carcinogenesis (hsa05204) and arachidonic acid metabolism (hsa00590). False discovery rate related to KEGG pathways is presented in Table 1.

### 3-3- Screening hub genes and modules from the PPI network.

The hub genes with Degree of interaction $\geq$ 9 are as followed: IGF1, PTGS2, PLG, HGF, ESR1,

CXCL12, CYP2B6, CYP2E1, SERPINE1, CCL2, CYP3A4, CYP2C9, EGR1, ALDH8A1, CYP1A2, CYP2C8, DCN, CXCL2, THBS1, CYP26A1, CYP2C19, CYP2A6, CYP4A11, PXR and IGFBP3. The hub genes with Betweenness centrality ≥ 100 are including PTGS2, IGF1, CYP2B6, ESR1, ALDH8A1, PLG, CYP2E1, CXCL12, DCN, SERPINE1, HGF, CYP3A4, CCL2, LUM and CXCL2. The hub genes with Eigenvector centrality ≥ 0.5 are including IGF1, PTGS2, PLG, HGF, ESR1, SERPINE1, CYP2B6, CCL2, CXCL12, EGR1, THBS1, CYP2C9, CYP3A4, CYP2E1, CYP1A2, CYP2C8 and CXCL2. The hub genes with the highest Closeness are including MT1X, MT1H, MT1E, MT1G, MT1F and MT1M. The list of 26 hub genes with the high centrality is presented in Supplementary Table 1.

Degree, Betweenness centrality, Closeness and Eigenvector centrality were visualized according to node color and size using Gephi software (Fig. 1, Supplementary Fig. 2). The hub genes were clustered into 7 modules using Gephi software (Fig. 2). Then, the genes associated with every module were deposited to Enrichr server, and GO analysis of the hub DEGs based on important modules was displayed in Table 2. Enrichment pathway analysis of the hub under-expressed genes with the highest Degree, Betweenness and Eigenvector in HCC was shown in Table 3.

### 3-4- Overall survival prediction

In order to predict overall survival of HCC patients, we deposited hub genes into GEPIA web server (http://gepia.cancer-pku.cn/). Among hub genes, nine genes were detected as prognostic genes in HCC including CYP2C9, ESR-1, CXCL2, CYP2C8, IGF-1, CYP3A4, CYP2E1, CERPINE-1 and PXR (Fig. 3).

### 4- Discussion

In the present study, IGF-1 exhibited the highest interaction degree, and IGF-1 expression level in the tumor tissue specimen of HCC patients was much lower than that in the liver tissue specimen of Hepatitis B patients. In agreement with the present study, Gao and colleagues identified IGF-1 as a hub gene that was downregulated in HCC [8]. Serum IGF-1 level is lower in the virus-associated HCC patients than that in HCC patients without viral infection [9-11]. In HCC patients, decreased serum IGF-1 level was associated with increased recurrence risk [12], advanced HCC and poorer overall survival [13-16]. In the present study, the growth hormone receptor (GHR) gene was also a downregulated DEG (adjusted p-value= 2.18E-06, Log FC= -2.00). In the patients with chronic liver disease such as HCC, decrease of hepatic responsiveness to GH is observed that results in low expression level of IGF-1 [10]. Furthermore, we detected that IGFBP-3 gene was significantly downregulated. IGFBP-3 acts as a reputed tumor suppressor gene and the principal regulator of IGF-1 bioactivity. Decreased expression level of IGFBP-3 has been associated with poor prognosis in patients with HCC [17].

The main functions of IGF-1 in liver include growth, differentiation and regeneration of hepatocytes [10, 18, 19]. According the recent studies, IGF-1 induced proliferation and epithelial-mesenchymal transition (EMT) of HCC cell lines [20, 21]. However, Adamek and colleagues in their study stated that systemic IGF-1 administration could ameliorate general liver function in HCC [10]. In addition to erythrocytosis and lymphocytopoiesis, IGF-1 has an important role in megakaryocyte differentiation and platelet production [22]. Moreover, pharmacological inhibition of the IGF-1R signaling reduced autophagy in *in vivo* models [23]. Autophagy acts as a tumor suppressor by maintaining genomic stability in normal cells [24], and it leads to death of HCC cells [25]. Based on aforementioned literature, low expression of IGF-1 gene is likely a biomarker of undifferentiated cells, adaptive immunosuppression, reduced autophagy, increased recurrence risk and advancement of HCC. According to Kaplan-Meier curve, IGF-1 has prognostic impact in the overall survival of HCC patients.

In the present study, expression level of plasminogen (PLG) in the tumor tissue specimen of HCC patients was significantly lower than that in the liver tissue specimen of Hepatitis B patients. According to prior literature, expression of PLG in HCC patients is downregulated [26, 27].

Plasmin is a potent inhibitor of angiogenesis [27, 28] and a serine protease in the fibrinolytic system [29]. Moreover, the removal of misfolded proteins and maintenance of tissue homeostasis are important physiological functions of plasmin [29]. In plasminogen-deficient mice, reduced growth and fertility, low survival as well as spontaneous fibrin deposition due to impaired thrombolysis is detected

[30]. Plasmin activity is required for tissue renewal [31], and loss of PLG impairs hepatocyte growth factor activation [32]. In addition, PLG mRNA levels are significantly associated with prolonged overall survival of cancerous patients [33]. On the other hand, plasmin is an activator of human monocytes, and it stimulates pro-inflammatory cytokines release [34], and promotes monocytes migration [35]. As a result, low expression of PLG gene is likely a biomarker of metastasis, innate immunosuppression, thrombosis and advancement of HCC.

In the present study the transcription factor early growth response 1 (EGR-1) was identified as the hub gene. EGR-1 expression increases during the process of growth, differentiation and regeneration [36-38]. Hao and colleagues reported that human tumor cell lines express little EGR-1 in contrast to their normal counterparts [39]. EGR-1 acts as a tumor suppressor in HCC via regulation of p53 and promoting apoptosis [40-42]. Besides, a recent study showed that EGR-1 knockdown disturbs fatty acid metabolism in breast cancer cell line [43]. Based on existing evidence, under-expressed EGR-1 might be a hallmark of undifferentiated tumor cells, decreased apoptosis and advanced HCC.

In the present study, the prostaglandin-endoperoxide synthase 2 (PTGS-2) or cyclooxygenase 2 (COX-2) was identified as the hub gene. This finding is consistent with the results of study published by Martín-Sanz and colleagues [44]. Several studies reported an extensive range of COX-2 expression in human HCC, from robust expression in differentiated tissue to absence of expression in undifferentiated area [45-48]. In partial hepatectomy, rapid expression of COX-2 in liver leads to inhibit apoptotic mechanisms [44, 49]. Moreover, overexpressed COX-2 promotes PI3K-dependent and extracellular signal-regulated kinases (ERKs) pathways [50] that accelerate cell proliferation and differentiation [39, 51, 52]. A recent study showed that overexpressed COX-2 in transgenic mice hepatocytes rises spontaneous HCC incidence [49]. On the other hand, a recent study demonstrated that COX-2 inhibition obviously repressed apoptosis, inflammation and prostaglandin E2 secretion in vascular endothelial cells [53]. According to aforementioned researches, under-expressed COX-2 might be related to undifferentiated liver cells, metastasis, immune suppression and HCC development, that warrants further experimental researches in future.

In the present study hepatocyte growth factor (HGF) was identified as the hub gene. In the adult, basal expression of HGF is important for normal tissue homeostasis [54]. HGF is responsible for proliferation and differentiation of hepatocytes, and repair of tissue injury [55-60]. Prior studies showed that HGF abrogated hepatic failure through anti-apoptotic mechanisms [61-63]. Shiota and colleagues reported that HGF is a negative growth regulator for HCC cells [64]. However, other studies reported that serum HGF level was significantly higher in patients with HCC than that in the control group [65] and HGF plays a critical role in HCC progression [60], migration of HCC cells [66, 67] and low survival time [59]. HGF is the regulator of inflammation and autoimmunity [68, 69] and limits the immune-mediated inflammation and liver damage [70]. As a result, low level of HGF expression might be associated with decreased differentiation of hepatocytes and chronic inflammation that warrants further experimental studies in future.

In the present study estrogen receptor alpha (ESR-1) was identified as the hub gene. This finding is in agreement with the published results of Sukocheva and colleagues [71]. According to earlier literature, estrogen receptor signaling provides protection against development of HCC [72, 73]. Decreased ESR-1 expression was significantly associated with liver damage, invasion and more size of tumor [74]. Some studies have indicated that 17β-estradiol treatment suppresses HCC cell proliferation and increases apoptosis in Hep3B cells [75]. In addition, the overexpression of ESR-1 induced apoptosis in Hep3B cells by increasing the expression of the TNF-α gene [76]. Decreased ESR-1 expression in the liver could contribute to increased gluconeogenesis and dyslipidemia [77]. As a result, reduced ESR-1 expression might be a hallmark of decreased apoptosis, metastasis and advanced HCC. According to Kaplan-Meier curve, ESR-1 has prognostic impact in the overall survival of HCC patients.

In the present study CYP1A2, 2A6, 3A4, 26A1, 2B6, 2C8, 2C9, 2C19, and 2E1, were identified as the hub genes. These enzymes are principally expressed in the liver and play a crucial role in the metabolism of antineoplastic agents [78]. In agreement with the present study, the numerous studies showed that the expression level of several CYP isoforms was decreased in HCC [26, 75, 79-84], that was associated with low tumor differentiation [85-87] increased risk of recurrence, advancement of HCC [75, 82, 88, 89] and poor prognosis [86-88, 90]. According to Kaplan-Meier curve, CYP2C8, CYP2C9, CYP2E1 and CYP3A4 have prognostic impacts in the overall survival of HCC patients.

In the present study, Aldehyde dehydrogenase 8 family member A1 (ALDH8A1/ ALDH12) was identified as the hub gene. ALDHs metabolize retinol to retinoic acid and regulate stem cell proliferation and differentiation [91, 92]. The defect in retinoid activity or responsiveness in HCC has been reported in some studies [93]. In a previous study, the administration of a synthetic retinoid reduced the post-therapeutic recurrence of HCC and improved the survival of patients with HCC [93]. In sum, the results of these studies show that low expression level of ALDH8A1 might be associated with less differentiation of cells, higher recurrence of HCC and poorer survival.

In the present study pregnane and xenobiotic receptor (PXR) or NR1I2 was identified as the hub gene. In agreement with the present study, Gillet and colleagues reported that in HCC Patients PXR gene is downregulated [94]. PXR lacks endogenous ligand. The role of nuclear receptor PXR in detoxification and clearance of xenobiotics has been confirmed [95, 96]. The downregulation of PXR reduces the hepatoprotective effects of autophagy [97]. PXR is important in liver regeneration by augmenting the proliferation of hepatocytes after a partial hepatectomy [98]. PXR regulates the apoptosis and cell proliferation in cancerous condition [99, 100]. Moreover, ligand-activated PXR regulates hepatic glucose and lipid metabolism and affects body metabolic homeostasis [101]. In chronic liver inflammation PXR activation reduced the levels of inflammatory cytokines in the liver [102, 103]. As a result, under-expression of PXR in HCC patients likely lead to persistent inflammation of the liver tissue, imbalance of glucose and lipid metabolism, disturbed metabolism of foreign substances and detoxification as well as reduced autophagy. According to Kaplan-Meier curve, PXR has prognostic impact in the overall survival of HCC patients.

In the present study, chemokines CCL2 (MCP-1), CXCL2 and CXCL12 were identified as the hub genes. Chemokines contribute to immune surveillance and chemotaxis regulation [104-106]. According to literature, CCL2 acts as a potent chemoattractant to recruit macrophages, tumor specific T cells, NK cells and neutrophils to the sites of inflammation [107-111]. Thus, the reduction of intra-tumor CCL2 contributes to the immune suppression in tumor tissues [107]. Moreover, CCL2 regulates T cell differentiation [112]. Marukawa and colleagues introduced immune-based gene therapy technique and the delivery of CCL2 gene as a novel strategy in the treatment of HCC patients [113]. Moreover, CXCL2 and CXCL12 expression is downregulated in HCC specimens [110, 114, 115]. CXCL2 inhibits HCC cell proliferation and promotes apoptosis in vitro [82, 114-116]. Moreover, B cells and T cells migration is regulated mainly by CXCL12 [104]. During chronic HBV infection, CXCL12 is a highly potent chemoattractant for liver-infiltrating lymphocytes [117]. Endogenous expression of CXCL12 by colonic carcinoma cells interrupts metastasis through induction of anoikis [118]. Semaan and colleagues reported that defect in CXCL12 expression has significant prognostic value [119]. In breast and colorectal carcinoma CXCL12 expression becomes silenced via DNA promoter hypermethylation, that enhances carcinoma progression [120, 121]. A previous study showed that increased cell differentiation enhances CXCL12 mRNA expression in HCT116 cells [122]. A prior study suggested that the increased CXCL2 may be a symptom of HCC occurrence in the future [109]. Collectively, it is possible that low expression of the chemokines might occur in undifferentiated hepatocytes and late stage of HCC and likely induces immunosuppression. According to Kaplan-Meier curve, CXCL2 has prognostic impact in the overall survival of HCC patients.

In the present study, Thrombospondin-1 (THBS-1/ TSP-1) was identified as the hub gene. THBS-1 is a member of family of extracellular matrix (ECM) proteins [123]. In agreement to the present study, Iida and colleagues found that THBS-1 level in normal tissue was more than that in HCC [124]. However, some studies reported that THBS-1 stimulates angiogenesis, invasiveness and advancement of HCC, and it is a significant prognostic factor in overall survival [125, 126]. THBS-1 stimulates platelet aggregation and its downregulation leads to attenuate hemostasis [127]. Some literature reported that THBS-1 acts as an angiogenesis inhibitor by stimulating endothelial cell apoptosis, inhibiting endothelial cell proliferation and migration [123, 128, 129]. A previous study indicated that in THBS-1-deficient mice with breast cancer, tumor proliferation was faster than in wild-type mice [130]. THBS-1 is often under-expressed during tumor development and its re-expression inhibits tumor growth via M1 macrophage recruitment and cytotoxicity against tumor cells [131]. THBS-1 promotes the inflammatory cytokine secretion through NF-κB signaling pathway [132]. It is possible that tumor cells during progression promote angiogenesis and inhibit innate immune surveillance in part via reducing THBS-1 expression.

In the present study plasminogen activator inhibitor-1 (SERPINE1/PAI-1) was identified as the hub gene. PAI-1 is a component of ECM. PAI-1 deficiency leads to MMP mediated collagen degradation as well as rapid degradation of fibrin [133, 134]. PAI-1 is a tumor suppressor [135]. PAI-1 prompts expression of pro-inflammatory cytokines and recruits peripheral monocytes to inflammation site [34]. Ren and colleagues reported that PAI-1 gene knockdown reduced TNF-α and IL-1β levels in cell supernatants and inhibited the NF-κB protein expression [136]. Tumor cells are enable to produce an immune tolerant microenvironment. Thus, applying novel therapeutic approaches for altering tumor microenvironment may participate in preventing HCC development. According to Kaplan-Meier curve, PAI-1 has prognostic impact in the overall survival of HCC patients.

Decorin (DCN) and lumican (LUM) are important components of the tumor microenvironment. In the present study, decorin and lumican were identified as the hub genes. Decorin is an extracellular anti-proliferative proteoglycan, and it evokes endothelial cell autophagy [137] and tumor cell apoptosis [138]. A recent study stated that level of decorin mRNA is downregulated in HCC [139]. Both decorin and lumican suppress tumorigenic growth and angiogenesis, and prevent metastasis in *in vitro* and *in vivo* tumor models [138-144]. Decorin enhances the expression of inflammatory factors such as TNF-α in macrophages [137] and inhibits TGF-β expression [145]. Moreover, lumican sustains adaptive antitumor immunity [143]. Based on a previous study, the expression of lumican decreased in advanced disease as compared to early stage of disease, and the tumor cell differentiation is largely responsible for lumican expression [146]. Recently Yang and colleagues have suggested that adenovirus expressing decorin could potentially be a candidate for the treatment of breast cancer metastasis [147]. As a result, decreased expression of decorin and lumican likely leads to reduced autophagy and apoptosis and increased angiogenesis and metastasis, as well as immunosuppression and poor outcome. Therefore, targeting matrix biosynthesis as a valuable therapeutic approach is worth paying attention in the future.

In conclusion, high throughput data analysis indicated that the genes involved in metabolism, differentiation, regeneration, coagulation regulation, immune system surveillance and ECM organization are under-expressed in advanced HCC. Furthermore, under-expression of several hub genes especially some CYP genes were with poor overall survival. The identification of the molecular mechanism of HBV carcinogenesis and important pathways repressed in HCC as well as the detection of prognostic proteins for designing novel therapeutic agents and molecular therapy is essential. Recently, some studies have discussed NK cell (innate immune) and T cell (adaptive immune) associated immunotherapies [148, 149], immune-based gene therapy [113] and cell therapy [150] as novel therapeutic paradigms for advanced HCC suppression. On the other hand, inhibition of hepatic inflammation may diminish the genomic instabilities that happen during hepatic necroinflammation and subsequent regeneration. So, attempt to timely identification of patients with chronic hepatitis B and prevent HCC occurrence through design of target-based anti-inflammatory agents is a rational approach.

## 5- Reference:

Figure 1

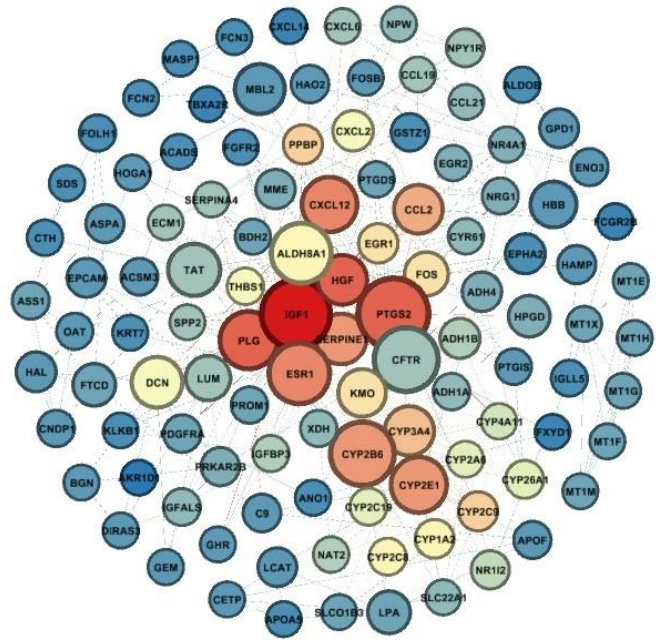

Figure 1: Representation of the centrality (Degree and Betweenness) of hub genes using Gephi software. The hub genes were defined as nodes. The red color shows the highest Degree of interaction, and the largest circle shows the highest Betweenness centrality.

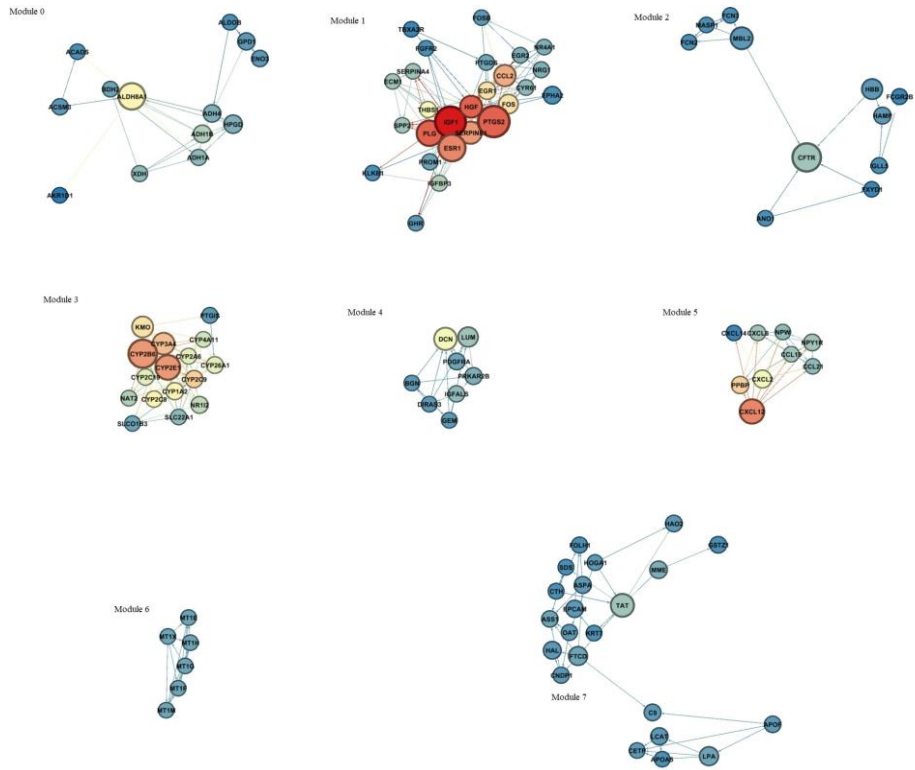

Figure 2: The modularity structures of hub DEGs in the PPI network using Gephi software. Thered nodes embody the highest Degree of interaction, and the larger node size embodies the higher Betweenness centrality. Owing to the presence of more hub genes in modules 1, 3, and 5 respectively, these three modules possess greater physiologic and therapeutic significance.

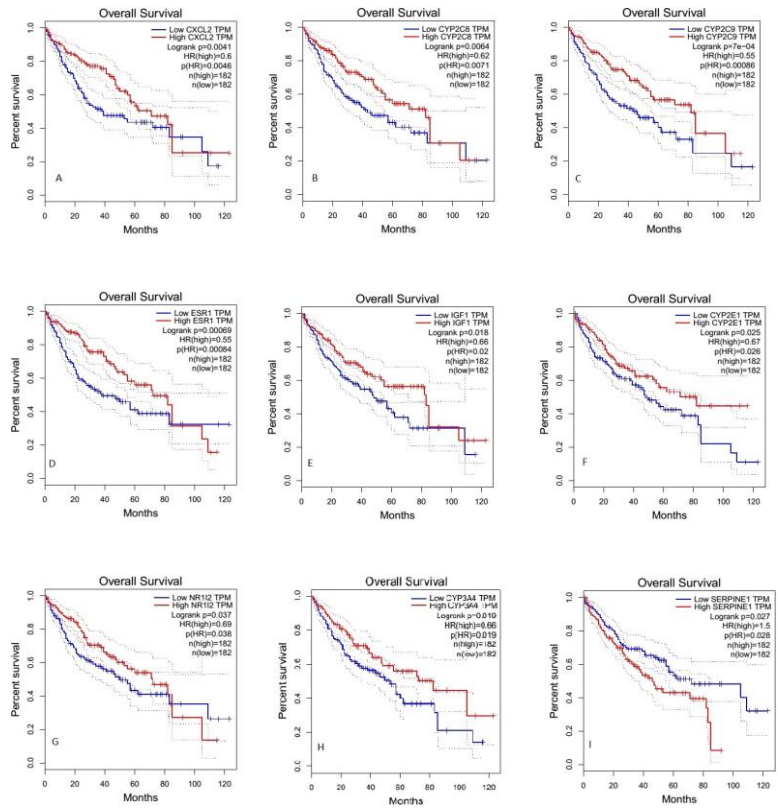

Figure 3: Prognostic value of hub genes. Based on GEPIA web server 9 prognostic genes in HCCare consisted of (A) CXCL2, (B) CYP2C8, (C) CYP2C9, (D) ESR-1, (E) IGF-1, (F) CYP2E1, (G) PXR, (H) CYP3A4, and (I) SERPINE-1. $p < 0.05$ was considered statistically significant.

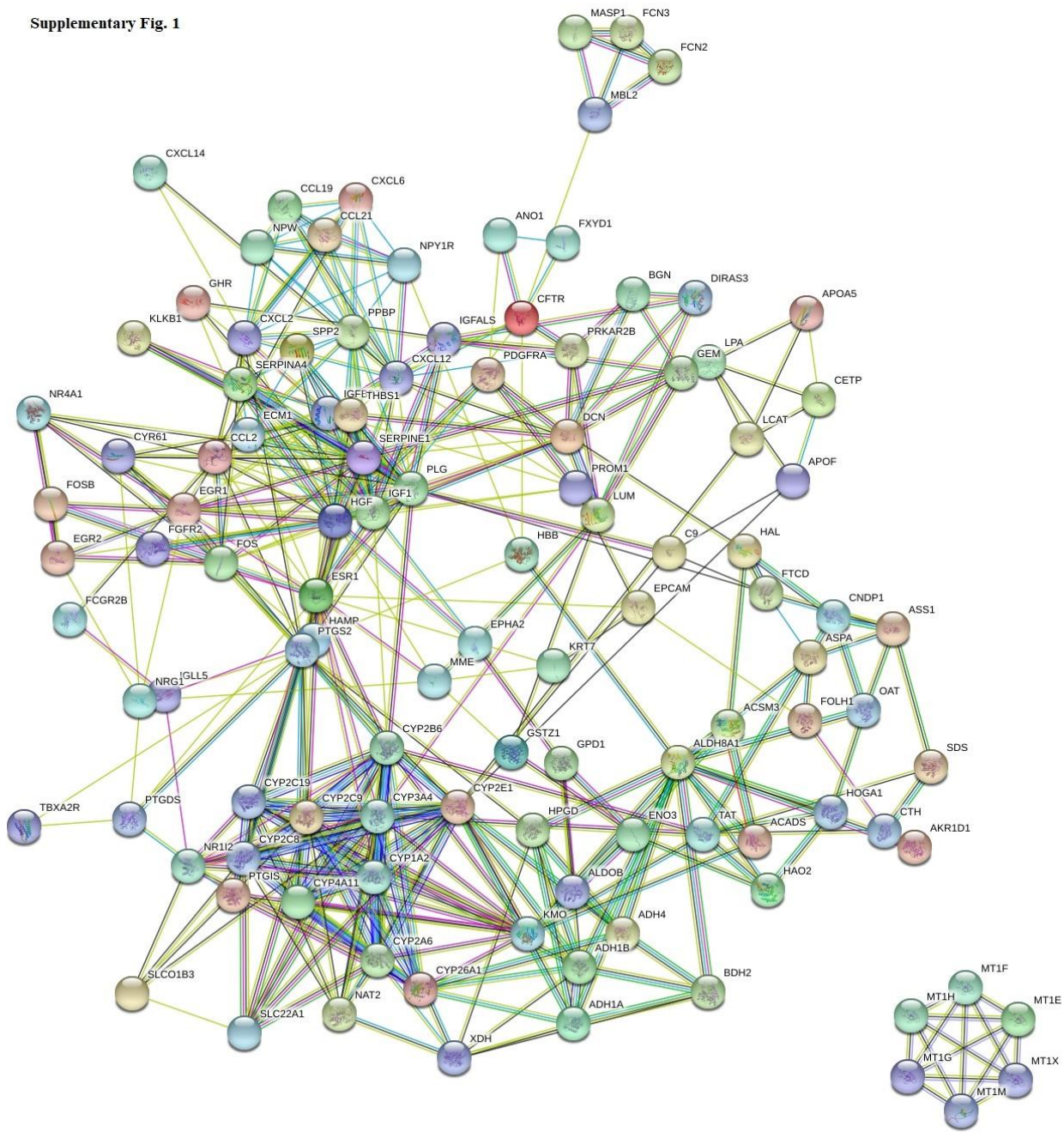

Supplementary Fig. 1: The protein-protein interaction (PPI) network of the under-expressed hub genes. The PPI network of the under-expressed genes was represented using the (STRING) (http://string-db.org/) database (Number of nodes=111, Number of edges=424, PPI enrichment p- value <1.0e-16).

**Supplementary Fig. 2**

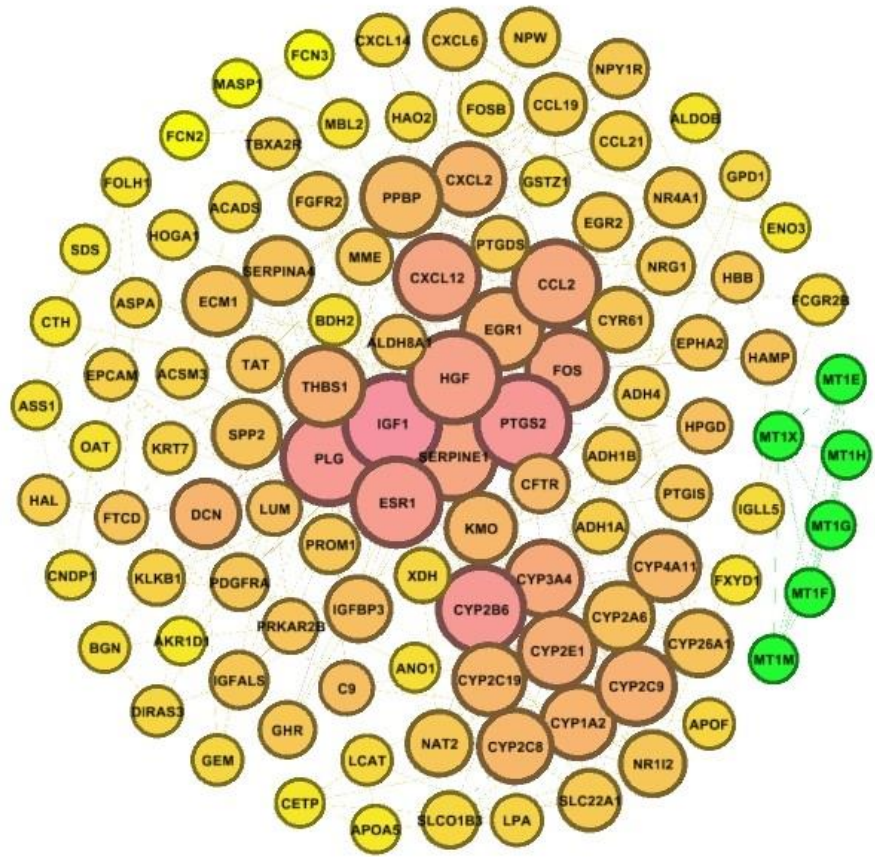

Supplementary Fig. 2: Representation of the centrality (Closeness and Eigenvector) of hub genes using Gephi software. The green color shows the highest Closeness, and the larger circles show the higher Eigenvector centrality.

**Table 1: KEGG pathways enriched by DEGs in STRING database**

| pathway ID | pathway description | gene count | matching proteins in network | False discovery rate (FDR) |
|---|---|---|---|---|
| 1100 | Metabolic pathways | 37 | ACADS,ACSM3,ADH1A,ADH1B,ADH4,AKR1D1,ALDOB,ASPA,ASS1,BDH2,CNDP1,CTH,CYP1A2,CYP2A6,CYP2B6,CYP2C19,CYP2C8,CYP2C9,CYP2E1,CYP3A4,CYP4A11,ENO3,FOLH1,FTCD,GSTZ1,HAL,HAO2,HOGA1,KMO,NAT2,OAT,PTGDS,PTGIS,PTGS2,SDS,TAT,XDH | 8.61E-17 |
| 5204 | Chemical carcinogenesis | 12 | ADH1A,ADH1B,ADH4,CYP1A2,CYP2A6,CYP2C19,CYP2C8,CYP2C9,CYP2E1,CYP3A4,NAT2,PTGS2 | 4.98E-13 |
| 830 | Retinol metabolism | 11 | ADH1A,ADH1B,ADH4,CYP1A2,CYP26A1,CYP2A6,CYP2B6,CYP2C8,CYP2C9,CYP3A4,CYP4A11 | 1.27E-12 |
| 982 | Drug metabolism - cytochrome P450 | 11 | ADH1A,ADH1B,ADH4,CYP1A2,CYP2A6,CYP2B6,CYP2C19,CYP2C8,CYP2C9,CYP2E1,CYP3A4 | 2.52E-12 |
| 4060 | Cytokine-cytokine receptor interaction | 11 | CCL19,CCL2,CCL21,CXCL12,CXCL14,CXCL2,CXCL6,GHR,HGF,PDGFRA,PPBP | 5.57E-06 |
| 590 | Arachidonic acid metabolism | 9 | CYP2B6,CYP2C19,CYP2C8,CYP2C9,CYP2E1,CYP4A11,PTGDS,PTGIS,PTGS2 | 2.31E-09 |
| 980 | Metabolism of xenobiotic by cytochrome P450 | 9 | ADH1A,ADH1B,ADH4,CYP1A2,CYP2A6,CYP2B6,CYP2C9,CYP2E1,CYP3A4 | 4.61E-09 |
| 4062 | Chemokine signaling pathway | 8 | CCL19,CCL2,CCL21,CXCL12,CXCL14,CXCL2,CXCL6,PPBP | 0.000111 |
| 4151 | PI3K-Akt signaling pathway | 8 | EPHA2,FGFR2,GHR,HGF,IGF1,NR4A1,PDGFRA,THBS1 | 0.00544 |
| 591 | Linoleic acid metabolism | 6 | CYP1A2,CYP2C19,CYP2C8,CYP2C9,CYP2E1,CYP3A4 | 2.73E-07 |
| 4978 | Mineral absorption | 6 | MT1E,MT1F,MT1G,MT1H,MT1M,MT1X | 7.39E-06 |
| 4610 | Complement and coagulation cascades | 6 | C9,KLKB1,MASP1,MBL2,PLG,SERPINE1 | 4.43E-05 |

Table 2. GO analysis of downregulated DEGs based on 3 important modularity classes

| Module | Biological process and molecular function | Differentially expressed genes (DEGs) | Combined score |
|---|---|---|---|
| 1 | regulated exocytosis (GO:0045055) | ECM, HGF, SERPINE1, PLG, IGF1, THBS1, SERPINA4 | 48.74 |
|  | phosphatidylinositol-4,5-bisphosphate 3-kinase activity (GO:0046934) | HGF, NRG1, ESR1, FGFR2 | 24.56 |
| 3 | exogenous drug catabolic process (GO:0042738) | CYP2C9, CYP2A6, CYP2C8, CYP2B6, CYP1A2, PXR, CYP2E1, CYP3A4, CYP2C19 | 114.44 |
|  | oxidoreductase activity (GO:0016712) | CYP2C9, CYP2A6, CYP2C8, CYP2B6, CYP1A2, CYP2E1, CYP2C19 | 86.33 |
| 5 | positive regulation of neutrophil chemotaxis (GO:0090023) | CXCL6, CCL21, CCL19, PPBP, CXCL2 | 63.88 |
|  | CXCR chemokine receptor binding (GO:0045236) | CXCL6, CXCL12, PPBP, CXCL2 | 58.88 |

Table 3. Pathway enrichment analysis of the hub under-expressed genes with interaction Degree ≥ 9 and Betweenness centrality ≥ 100 in HCC.

| Down-expressed hub genes | Repressed pathways | Combined score | Modularity class |
|---|---|---|---|
| ALDH8A1 | RA biosynthesis pathway (R-HSA-5365859) | 27.63 | 0 |
| IGF1 | PI3K-Akt signaling pathway (hsa04151) | 31.92 | 1 |
| | p53 signaling pathway (hsa04115) | 22.95 | |
| | Platelet degranulation (R-HSA-114608) | 54.03 | |
| PTGS2 | Synthesis of Prostaglandins and Thromboxanes (R-HSA-2162123) | 15.09 | 1 |
| | Integrated Pancreatic Cancer Pathway (WP2377) | 35.85 | |
| PLG | Complement and coagulation cascades (hsa04610) | 14.40 | 1 |
| | Platelet degranulation (R-HSA-114608) | 54.03 | |
| HGF | PI3K-Akt signaling pathway (hsa04151) | 31.92 | 1 |
| | Platelet degranulation (R-HSA-114608) | 54.03 | |
| ESR1 | Proteoglycans in cancer (hsa05205) | 17.09 | 1 |
| | Signaling by ERBB4 (R-HSA-1236394) | 24.58 | |
| | Integrated Pancreatic Cancer Pathway (WP2377) | 35.85 | |
| SERPINE1 | p53 signaling pathway (hsa04115) | 22.95 | 1 |
| | Platelet degranulation (R-HSA-114608) | 54.03 | |
| | Oncostatin M Signaling Pathway (WP2374) | 36.26 | |
| CCL2 | Signal Transduction (R-HSA-162582) | 19.07 | 1 |
| | Oncostatin M Signaling Pathway (WP2374) | 36.26 | |
| | Tamoxifen metabolism (WP691) | 79.21 | |
| IGFBP3 | p53 signaling pathway (hsa04115) | 22.95 | 1 |
| | Regulation of IGF-1 (R-HSA-381426) | 24.56 | |
| THBS1 | PI3K-Akt signaling pathway (hsa04151) | 31.92 | 1 |
| | p53 signaling pathway (hsa04115) | 22.95 | |
| | Rap1 signaling pathway (hsa04015) | 22.72 | |
| | Platelet degranulation (R-HSA-114608) | 54.03 | |
| EGR1 | AGE-RAGE signaling pathway (hsa04933) | 14.87 | 1 |
| | Cytokine Signaling in Immune system (R-HSA-1280215) | 15.09 | |
| | Oncostatin M Signaling Pathway (WP2374) | 36.26 | |
| CYP2B6, CYP2E1, CYP3A4, CYP2C9, CYP1A2, CYP2C8, CYP26A1, CYP2C19, CYP2A6, CYP4A11 | Retinol metabolism (hsa00830) | **68.27** | **3** |
| | Drug metabolism (hsa00982) | **67.33** | |
| | Arachidonic acid metabolism (hsa00590) | **59.45** | |
| | Chemical carcinogenesis (hsa05204) | **56.95** | |
| PXR | Nuclear Receptor transcription pathway (R-HSA-383280) | **4.32** | **3** |
| | Nuclear Receptors in Lipid Metabolism (WP299) | **83.57** | |
| DCN, LUM | Proteoglycans in cancer ( hsa05205) | **11.86** | **4** |
| | glycosaminoglycan metabolism ( R-HSA-3560782) | **33.20** | |
| CXCL2, CXCL12 | Chemokine signaling pathway ( hsa04062) | **56.40** | **5** |
| | G alpha (i) signalling events ( R-HSA-418594) | **71.91** | |

**Supplementary Table 1. The down-regulated hub DEGs in HCC**

| Gene symbol | Log FC | Adjusted p-value | Gene symbol | Log FC | Adjusted p-value |
|---|---|---|---|---|---|
| **CYP1A2** | -4.44 | 2.67E-10 | **THBS1** | -2.25 | 6.84E-08 |
| **CYP26A1** | -3.78 | 7.99E-13 | **CYP2A6** | -2.25 | 2.62E-05 |
| **DCN** | -3.17 | 4.26E-08 | **CXCL2** | -2.12 | 3.09E-07 |
| **CXCL12** | -3.12 | 7.73E-10 | **ESR1** | -2.1 | 1.75E-08 |
| **HGF** | -2.9 | 4.21E-08 | **PXR** | -1.94 | 8.16E-06 |
| **LUM** | -2.73 | 0.0042 | **IGF1** | -1.86 | 4.29E-06 |
| **CYP2B6** | -2.5 | 3.75E-08 | **CYP2E1** | -1.8 | 0.00019 |
| **IGFBP3** | -2.5 | 1.70E-12 | **PLG** | -1.78 | 6.65E-13 |
| **CYP3A4** | -2.49 | 4.11E-06 | **CCL2** | -1.73 | 0.00024 |
| **EGR1** | -2.42 | 3.92E-11 | **CYP2C9** | -1.71 | 0.00049 |
| **SERPINE1** | -2.39 | 3.05E-07 | **CYP2C19** | -1.67 | 3.23E-08 |
| **PTGS2** | -2.29 | 2.97E-05 | **CYP2C8** | -1.62 | 0.00036 |
| **CYP4A11** | -2.29 | 3.17E-09 | **ALDH8A1** | -1.56 | 1.82E-07 |